\documentclass[runningheads]{llncs}
\usepackage[square,numbers,sort&compress]{natbib}
\usepackage{booktabs,bbm,amsmath,tabularx}
\usepackage{xcolor,bm,amsmath,graphicx,comment,bigints}
\usepackage{algorithm,algorithmic,url}
\usepackage{subcaption}
\begin{document}
\title{Community Detection in Medical Image Datasets: Using Wavelets and Spectral Methods}
\titlerunning{Community Detection in Medical Image Datasets}
%
\author{Roozbeh Yousefzadeh}
%
%
\institute{Yale Center for Medical Informatics and VA CT Healthcare System}
%
\maketitle              

\vspace{-.5cm}

\begin{abstract}
Medical image datasets can have large number of images representing patients with different health conditions and various disease severity. When dealing with raw unlabeled image datasets, the large number of samples often makes it hard for experts and non-experts to understand the variety of images present in a dataset. Supervised learning methods rely on labeled images which requires a considerable effort by medical experts to first understand the communities of images present in the data and then labeling the images. Here, we propose an algorithm to facilitate the automatic identification of communities in medical image datasets. We further demonstrate that such analysis can be insightful in a supervised setting when the images are already labeled. Such insights are useful because, health and disease severity can be considered a continuous spectrum, and within each class, there usually are finer communities worthy of investigation, especially when they have similarities to communities in other classes. In our approach, we use wavelet decomposition of images in tandem with spectral methods. We show that the eigenvalues of a graph Laplacian can reveal the number of notable communities in an image dataset. Moreover, analyzing the similarities may be used to infer a spectrum representing the severity of the disease. In our experiments, we use a dataset of images labeled with different conditions for COVID patients. We detect 25 communities in the dataset and then observe that only 6 of those communities contain patients with pneumonia. We also investigate the contents of a colorectal cancer histology dataset.

\end{abstract}
%
%
%


\section{Introduction}

\vspace{-.2cm}

Analyzing the contents of medical image datasets is not a straightforward task. In practice, it is useful to label images based on health or severity of the disease. Although health and disease can be considered a continuous spectrum, for practical purposes, we usually need to divide that spectrum into specific groups/labels. For example, in the case of analyzing chest X-ray images with respect to the COVID-19 disease, it is useful to define labels such as healthy, mild, severe, and pneumonia. This is not motivated by machine learning, rather by different category of medical procedures that should follow.

{\bf Dealing with unlabeled image datasets.} Annotating and labeling medical images requires medical expertise and it is an expensive procedure, prone to mistakes and noisy labels \citep{zhang2020collaborative}. This makes it sometimes prohibitive to create and analyze large medical image datasets for machine learning and automation. Despite all these difficulties, medical institutions have plenty of raw medical images available, and automating the process of analyzing such datasets and identifying groups of similar images can be beneficial for two reasons. 

First, such analysis provides insights about the variety of images present in a raw dataset. For example, if we gather the chest X-ray images of all COVID-19 patients in a given hospital at a given day, it would not be clear how much variety will be present in the gathered data. Some X-rays might be from the side of the patients, some might be from their front. Some images might correspond to patients who have developed different degrees of pneumonia. Each of these types can be considered a specific community in the dataset. It would be useful to know how many communities of similar images are present in a dataset. We show that eigenvalue analysis of a graph Laplacian can provide an estimate of the number of such communities.

Second, automatically detecting groups of similar images can facilitate the labeling process, because the medical expert can then review the groups of images, instead of going through all the images one by one. Here, we show that wavelet decomposition of images in tandem with clustering can extract the groups of similar images from the data.

{\bf Dealing with labeled image datasets.} After a medical image dataset is labeled, or when we are given a labeled image dataset, it would be useful to analyze the similarities within each class, and also analyze the cross-class similarities. Mistakes in labeling is not unusual, even by experts, especially when dealing with large datasets. Analyzing the similarities may be able to identify such mistakes. An image that is isolated and dissimilar from other images in a class might actually be a mislabeled image; and even when when such images are correctly labeled, it would still be useful to be informed about their existence, and understand the reason behind their dissimilarity to other images of the class. In fact, identifying dissimilar images of the same class are useful for efficient training of models, e.g., triplet mining \citep{hermans2017defense,yuan2020defense}. Moreover, analyzing the similarities of labeled images may help us infer a disease spectrum representing the severity of disease among patients as we discuss in our results.










{\bf Related work.} There are a few studies that have used community detection methods to detect specific items in images \citep{linares2017segmentation,javed2020cellular}. In those approaches, each community consists of certain pixels inside an image and not a group of similar images inside a dataset. \citet{trivizakis2021neural} used wavelets to extract features from images, and then, used those features to train a classification model on histology images of colorectal cancer. This shows the effectiveness of wavelets in extracting features. However, their method is not comparable to ours as their focus is on training a classification model on a labeled dataset, not identifying groups of similar images, analyzing in-class and cross-class similarities, and inferring a disease spectrum, as we do.

There is a rich literature on community detection algorithms for tabular data and networks \citep{li2018local,shi2019locally}, but those methods are not readily applicable to image datasets. In the object recognition literature, there are methods that create an embedding for images, but their computational method significantly differs from ours. First, they are not for medical image datasets, rather for object recognition. Second, they are not concerned with detecting communities of similar images in the datasets. Third, they do not use spectral methods to analyze the abundance of similarities. Fourth, they compare images either by solving expensive optimization problems \citep{vo2019unsupervised}, or by comparing image representations in an inner layer of a trained deep network \citep{birodkar2019semantic}. This last approach requires a trained model in the first place which can be very expensive.

Recently, \citet{das2019efficient} suggested a method to identify images in a medical image datasets most similar to a specific query image. This method is specifically designed for histology images of breast. It uses wavelets to identify specific patches in images, and eventually trains a convolutional neural network and uses the representations learned by the model to identify similar images to the query image. Although this method has similarities to our method, it also has considerable differences. First, its requires training a neural network on images. Second, it is specific to histology images of breast and detection of mitotic cells. Third, it only identifies images similar to a single query image, and does not analyze the similarities in the entire dataset while we perform that task by forming a graph Laplacian for entire datasets and analyzing its eigenvalues. \citet{mazzei2021unsupervised} proposes an unsupervised approach


\vspace{-.5cm}

\section{Our Algorithm}

\vspace{-.3cm}

Our algorithm relies on wavelets, spectral methods, and clustering.



{\bf Wavelets.} Wavelets are a class of functions and one of the most capable tools to systematically process images and extract features and patterns from them. The difficulty of working with images and many signals arises from the spatial complexity of patterns and structures in them. What makes an X-ray image to represent signs of pneumonia cannot be explained by one or even a few pixels, rather, it may be explained by the specific patterns that appear in various regions of an image.

Wavelets were developed building on the scientific knowledge of Fourier transform in the context of image and signal processing. Notably, \cite{daubechies1990wavelet} showed that wavelets perform better than windowed Fourier transform on visual signals, because wavelets handle the frequencies in a nonlinear way. The family of Daubechies wavelets \citep{daubechies1992ten} are one of the most successful types of wavelet transformation, and we use them in this paper. The orthogonality of Daubechies wavelets is particularly useful for feature extraction, because orthogonality in this setting implies the filters are independent and each filter is measuring a specific feature in the image signals. To process images with wavelets, we use the function
\begin{equation}
    [\omega,\beta] = \mathrm{wavedec}(x,\Omega,N),
\end{equation}
which takes as input an image $x$, a wavelet basis $\Omega$, and level number $N$. It returns a vector of real numbers $\omega$, representing the wavelet coefficients obtained from convolving $x$ with $\Omega$, and a book keeping matrix $\beta$ containing the dimensions of wavelet coefficients by level. This operation is reversible, therefore, given $\omega$, $\beta$, and $\Omega$, we can return to pixel space and reconstruct the image $x$, which we denote its operation by
\begin{equation}
    x = \mathrm{waverec}(\omega,\beta,\Omega).
\end{equation}
For a given $N$, $\beta$ will be constant for all images of the same size.




{\bf Radiomics.} Radiomics refers to an emerging class of computational methods that aim to extract features from medical images that can be useful for clinical decision making and outcome prediction \cite{gillies2016radiomics,rizzo2018radiomics}. 
In certain applications, these methods have been able to extract, from images, features that are not easily detectable by eye, and they have been able to characterize clinically useful phenotypes. For example, \citet{lafata2019exploratory} identified 39 radiomic features from CT images of lungs as potential biomarkers for pulmonary function. Computing the radiomic features usually relies on statistical methods and consider the shapes, intensities, textures of items in the images \cite{rizzo2018radiomics}. Intensity related features are usually computed using first-order statistics, and texture related features are computed using second-order statistics. In a few occasions, more sophisticated methods such as wavelets are used to extract radiomic features, e.g., \cite{aerts2014decoding}.





{\bf Our Algorithm.} Our algorithm first decomposes all images using a wavelet basis $\Omega$, and organize them in a matrix $\mathcal{C}$ with rows corresponding to images and columns corresponding to wavelet coefficients (lines~1 to~4). About the choice of $\Omega$ we have observed empirically that Daubechies-1, 2, and 3, work well in extracting features from images. Higher level wavelets can extract some extra features corresponding to finer details in images, but those finer details are not always useful for analyzing the similarities of images. In fact, when we use higher level wavelets, like Daubechies-5, and extract more features from images, the feature selection part of our algorithm discards those extra features. For each dataset, we recommend a few different wavelet bases to be tested, starting from Daubechies-1. Overall, the choice of $\Omega$ does not affect our empirical results.

Next, our algorithm selects a subset of wavelet coefficients according to their Laplacian score \citep{he2005laplacian} and using the function $\mathrm{fsulaplacian}(.)$ (line~5). Laplacian score is a feature selection method based on Laplacian eigenmaps and Locality Preserving Projection \citep{he2004locality}, specifically designed for unsupervised settings. Wavelet coefficients with scores less than the threshold $\tau$ will be discarded (line~6). Feature selection with Laplacian score is a standard method and there are standard recommendations for the choice of $\tau$. We recommend several values to be tested for $\tau$ to make sure useful features are not discarded. It is possible to used other feature selection methods as well. In the past, we have used rank-revealing QR factorization \citep{chan1987rank}, but we prefer the Laplacian score because it relates to later steps of our algorithm where we derive the graph Laplacian.

Our algorithm then computes a distance matrix, $D$, by applying the function $\mathrm{pdist}(.)$ on $\mathcal{C}$ (line~7). $\mathrm{pdist}(.)$ measures the pairwise distances between the rows of $\mathcal{C}$ and returns a symmetric square matrix $D$. To measure the distances, we use the distance metric $\mathcal{M}$. In practice, we have found the {\em correlation distance} to be an effective metric. Other metrics such as cosine similarity may work as well. We then convolve the $D$ with a Gaussian kernel to turn it into an affinity matrix, $W$ (line~8). In this line, $std(.)$ returns the standard deviation, $exp(.)$ is the exponential function, and $\odot$ is the Hadamard product. The diagonal elements of $W$ are set to zero. Using the affinity matrix and the function $\mathrm{glaplacian}(.)$, we derive the graph Laplacian of the data (line~9). The eigenvalues of the graph Laplacian will let us identify the number of clusters in the data, $n_c$ (lines~10-11). This is a standard method suggested by \citet{von2007tutorial}. To estimate the number of clusters, it is possible to use alternative methods as well.

\vspace{-.6cm}

\begin{algorithm}[h]
\caption{Wavelet Spectral Decomposition for Community Detection (WSDCD): Algorithm for detecting communities of similar images in datasets
}
\label{alg:main}
\textbf{Inputs}: Dataset of images $\mathcal{P}$, wavelet basis $\Omega$, distance metric $\mathcal{M}$, feature selection threshold $\tau_w$, eigenvalue threshold $\tau_c$\\
\textbf{Outputs}: Communities in the dataset $idc$\\
\vspace{-.3cm}
\begin{algorithmic}[1] 
\STATE Count total number of images in $\mathcal{P}$ as $n$
\FOR{$i=1$ to $n$}
    \STATE $\mathcal{C}(j,:,i) = \mathrm{wavedec}(\mathcal{P}\{i\},\Omega)$
\ENDFOR
\vspace{.1cm}
\STATE $[idw,scorew] = \mathrm{fsulaplacian}(\mathcal{C})$
\vspace{.1cm}
\STATE $\mathcal{C}(:,idw(scorew < \tau_w)) = []$
\vspace{.1cm}
\STATE $D = \mathrm{pdist}(\mathcal{C},\mathcal{M})$
\vspace{.1cm}
\STATE $W = \frac{1}{\mathrm{std}(S)} \; \mathrm{exp}(S\odot S)$
\vspace{.1cm}
\STATE $L = \mathrm{glaplacian}(W)$
\vspace{.1cm}
\STATE $\lambda = \mathrm{eig}(L)$
\vspace{.1cm}
\STATE estimate the number of clusters, $n_c$, based on the eigen-gaps
\vspace{.1cm}
\STATE $idc = \mathrm{cluster}(\mathcal{C},\mathcal{M})$
\STATE \textbf{return} $idc$
\end{algorithmic}
\end{algorithm}

\vspace{-.4cm}

Finally, we cluster the images into $n_c$ clusters based on their affinities captured in $W$ and using a clustering function of choice (line~12). As a result, similar images will appear in each of the clusters and we will be able to provide them to medical experts for further analysis.

\vspace{-.4cm}

\section{Results}

\vspace{-.4cm}



\begin{figure}[h]
  \centering
     \begin{subfigure}[b]{0.25\textwidth}
         \centering
         \includegraphics[width=1\linewidth]{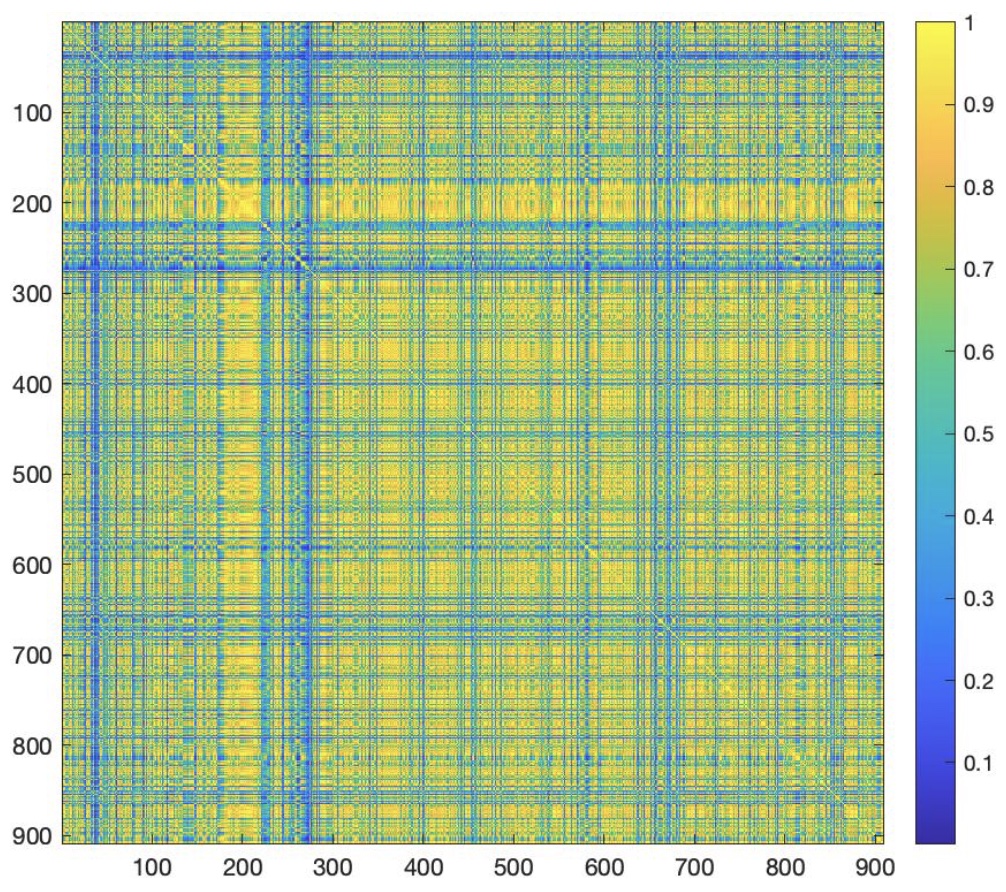}
         \caption{}
         \label{fig:sim_matrix}
     \end{subfigure}
     \begin{subfigure}[b]{0.55\textwidth}
         \centering
         \includegraphics[width=1\linewidth]{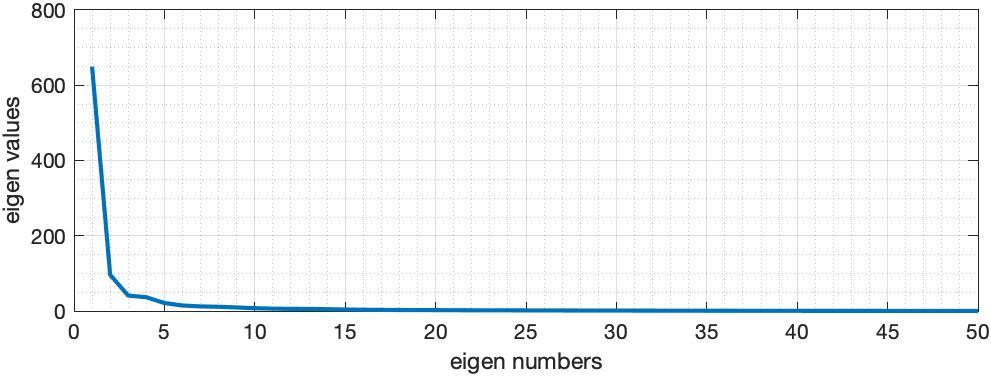}
         \caption{}
         \label{fig:eigs}
     \end{subfigure}      
  \caption{Similarity matrix obtained based on wavelet decomposition of all images in the COVID-19 dataset and the distribution of its eigenvalues.}
  \label{fig:sim_matrix_eigs}
\end{figure}

{\bf Dataset on COVID-19 Radiology.} We use the dataset provided by \citet{cohen2020covid} which contains a mixture of 909 chest X-ray and CT-scan images of patients diagnosed with COVID-19. These images correspond to patients with various degrees of the disease. We proceed with analyzing the dataset by first decomposing the images with Daubechies-3 wavelets \citep{daubechies1992ten}. We then measure the cosine similarity of wavelet coefficients of the images. Figure~\ref{fig:sim_matrix} shows the similarity matrix obtained from this analysis.


Using the similarity matrix, we then compute its normalized graph Laplacian. Figure~\ref{fig:eigs} shows the eigenvalues of the Laplacian. As we can see, the number of large eigenvalues are not many. In fact, the eigenvalues beyond the 25\textsuperscript{th} are very close to zero. Based on this, we choose the number of clusters (i.e., image communities) as 25, and proceed with spectral clustering of the images.

\vspace{-.3cm}

\begin{figure}[h]
  \centering
     \begin{subfigure}[b]{0.4\textwidth}
         \centering
         \includegraphics[width=1\linewidth]{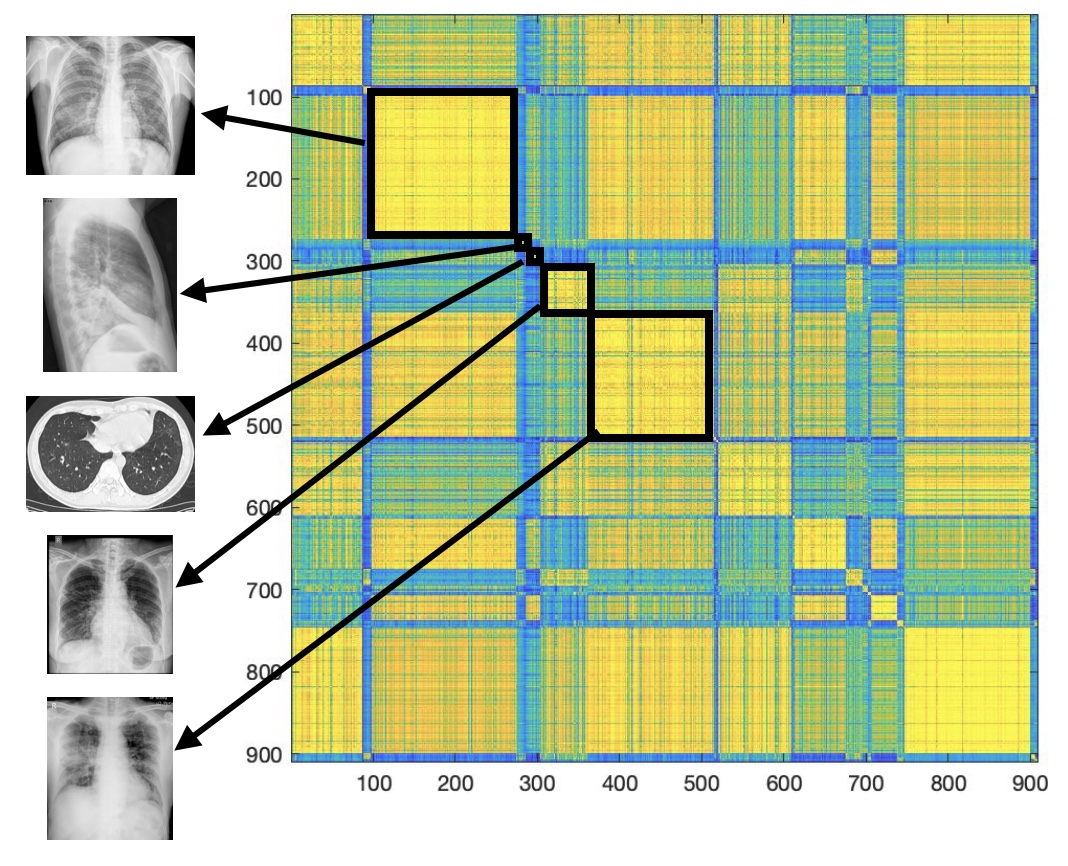}
         \caption{}
         \label{fig:clusters_all}
     \end{subfigure}
     \begin{subfigure}[b]{0.45\textwidth}
         \centering
         \includegraphics[width=1\linewidth]{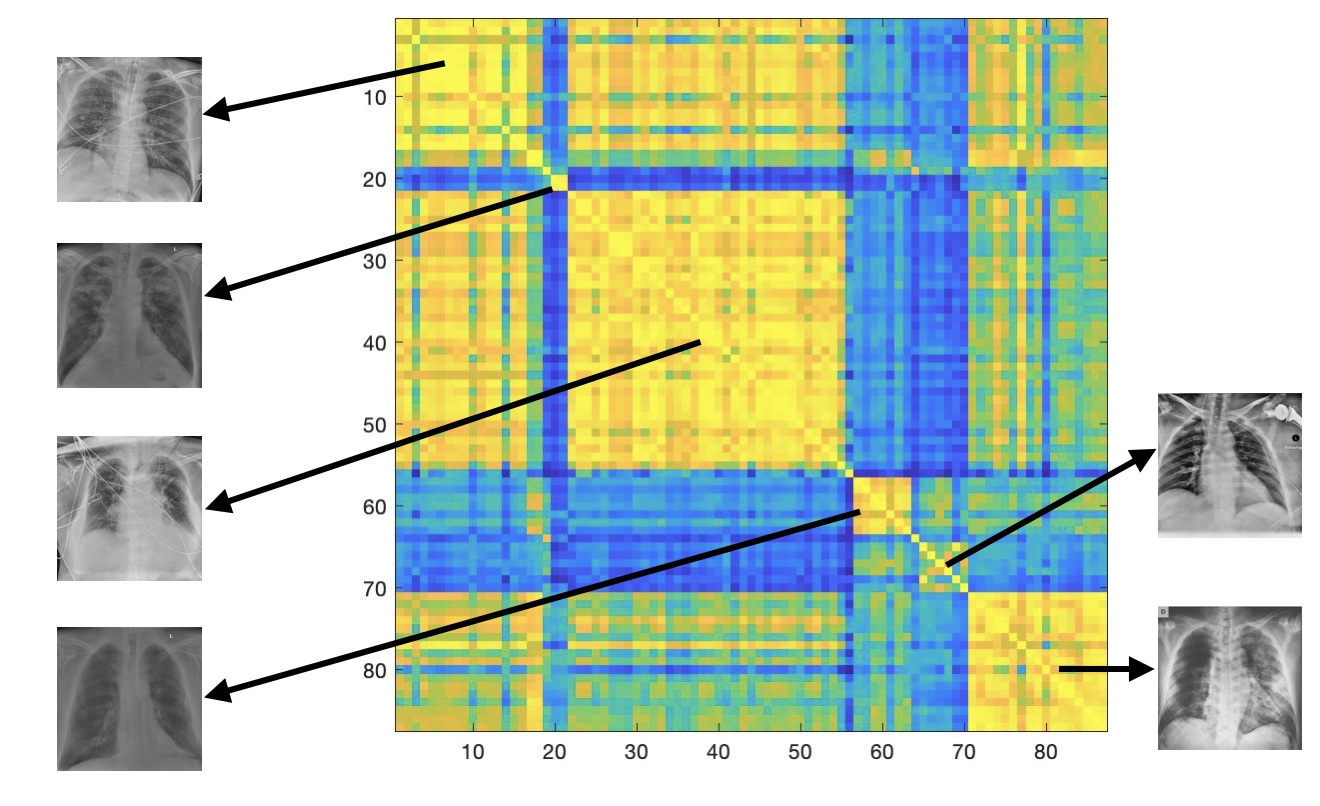}
         \caption{}
         \label{fig:clusters_pn}
     \end{subfigure}
  \caption{\textbf{(a)} Images of different mode appear in different clusters, so does images with different severity of disease. \textbf{(b)} Subset of similarity matrix for images annotated with pneumonia.}
  \label{fig:clusters}
\end{figure}

Figure~\ref{fig:clusters_all} shows the same similarity matrix as in Figure~\ref{fig:sim_matrix} after re-ordering the rows and columns of the matrix based on the appearance of images in the clusters. Each block along the diagonal of the matrix corresponds to one of the clusters in our image dataset. The off-diagonal blocks reveal the similarity of clusters with each other.



\vspace{-.3cm}

Further examination of these clusters reveal that patients with pneumonia appear only in 6 of the 25 clusters, as shown in Figure~\ref{fig:clusters_pn}. We also see that images of different modality appear in separate clusters, as shown in Figure~\ref{fig:clusters_all}. In practice, when we receive such dataset with no labels, our algorithm can detect the communities of similar images. This would facilitate the process of labeling images by medical experts. Later in Section~\ref{sec:spectrum}, we will show how analyzing the similarities can be used to infer a disease spectrum representing its severity.

\begin{figure}[h]
  \centering
     \begin{subfigure}[b]{0.33\textwidth}
         \centering
         \includegraphics[width=1\linewidth]{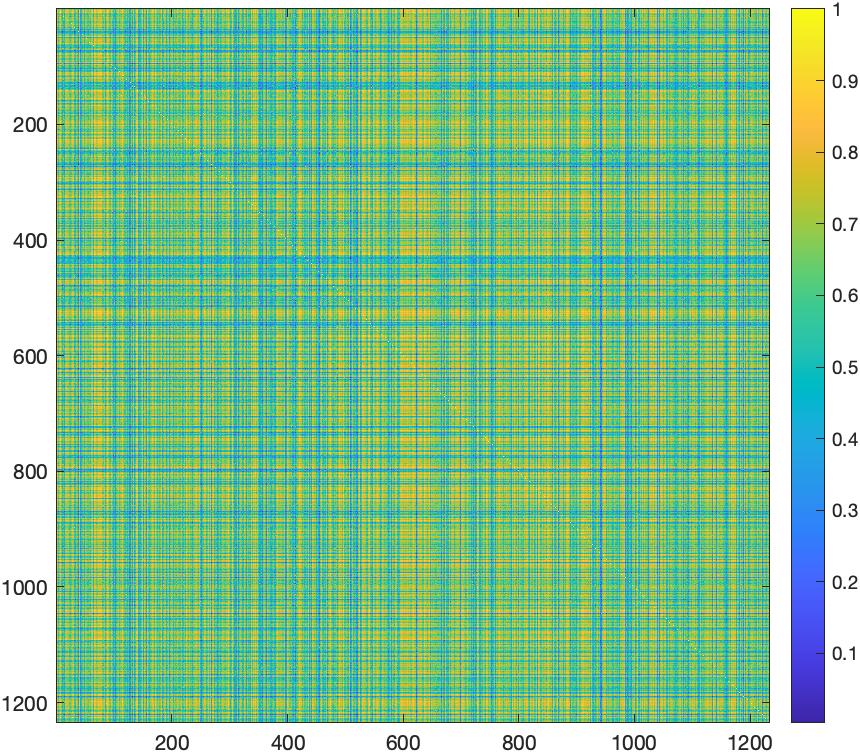}
         \caption{}
         \label{fig:crc_9_similarity_orig}
     \end{subfigure}
     \begin{subfigure}[b]{0.45\textwidth}
         \centering
         \includegraphics[width=1\linewidth]{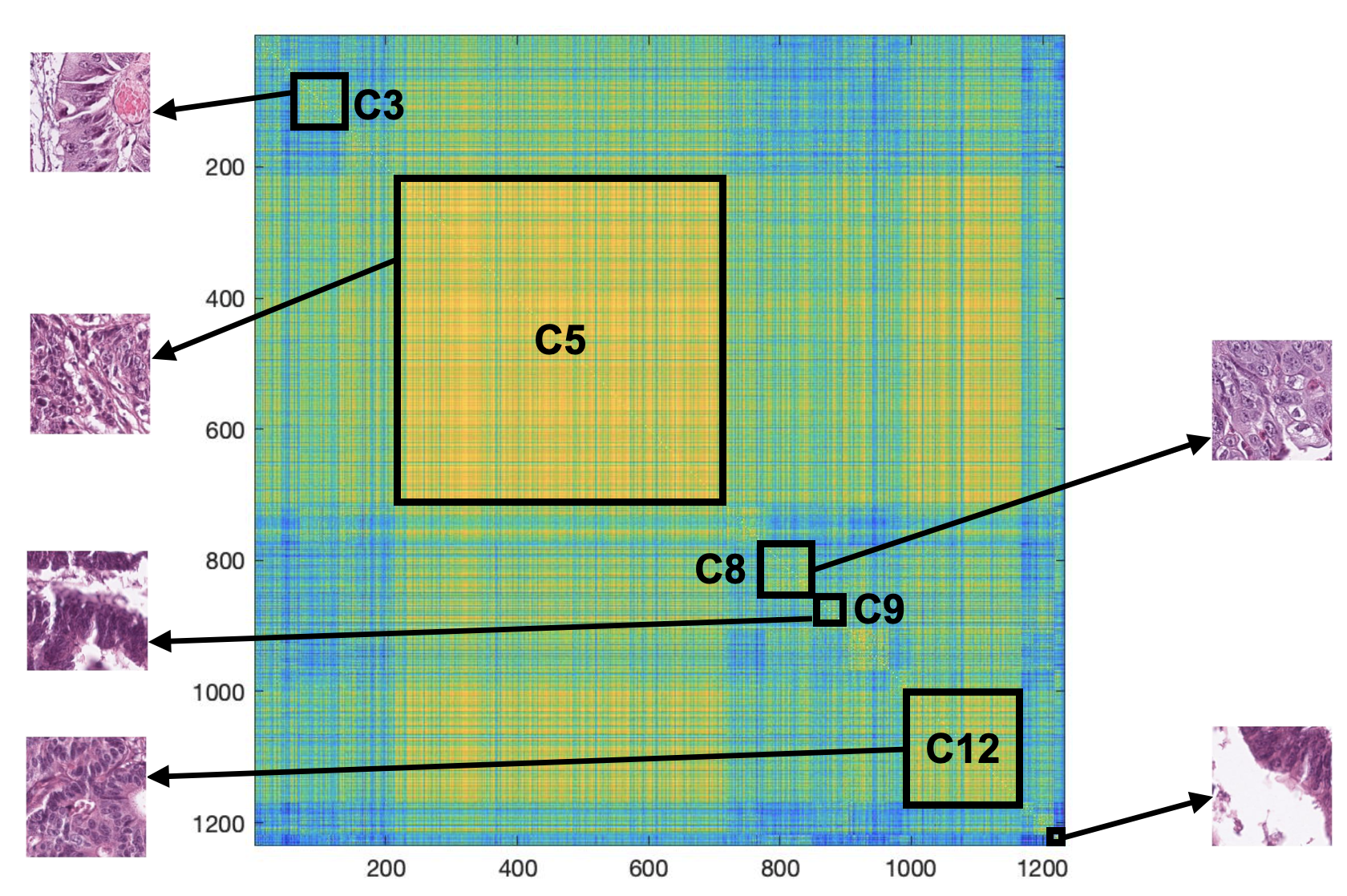}
         \caption{}
         \label{fig:crc_9_similarity_reordered}
     \end{subfigure}
     \caption{\textbf{(a)} Similarity matrix for the Colorectal Adenocarcinoma Epithelium class. \textbf{(b)} Reordered similarity matrix based on the clusters in the data along with samples from some of the clusters. Note that all the clusters are labeled as one malignant type of cancerous tissue.}
  \label{fig:crc_1_similarity}
\end{figure}

{\bf Colorectal Adenocarcinoma Epithelium (TUM) tissue.} Here, we study the last class of the colorectal cancer (CRC) histological image dataset \citep{kather2019predicting}. This class contains 1,233 images of Colorectal Adenocarcinoma Epithelium (TUM) tissue. Figure~\ref{fig:crc_9_similarity_orig} shows the resulting similarity matrix demonstrating a relatively larger variety among images. Using the eigenvalues of graph Laplacian, we choose the number of clusters to be 15. Figure~\ref{fig:crc_9_similarity_reordered} shows the reordered similarity matrix after the clustering and also samples from each of the cluster.

In Figure~\ref{fig:crc_9_similarity_reordered}, note that all images in all the clusters are considered one malignant type of cancerous tissue. But, there are still different varieties in their patterns. Each cluster appears as a diagonal block in the similarity matrix. By looking at the off-diagonal blocks of the matrix, we can identify which clusters are more similar to each other and which clusters are less similar. For example, note that cluster C3 is more similar to cluster C5 compared to other clusters. Our algorithm quickly reveals the variety of images present in this class which can help us to further divide the class into sub-classes for more detailed study.



\section{Inferring the disease spectrum by analyzing in-class and out-class similarities} \label{sec:spectrum}

\begin{figure}[h]
  \centering
   \includegraphics[width=.8\linewidth]{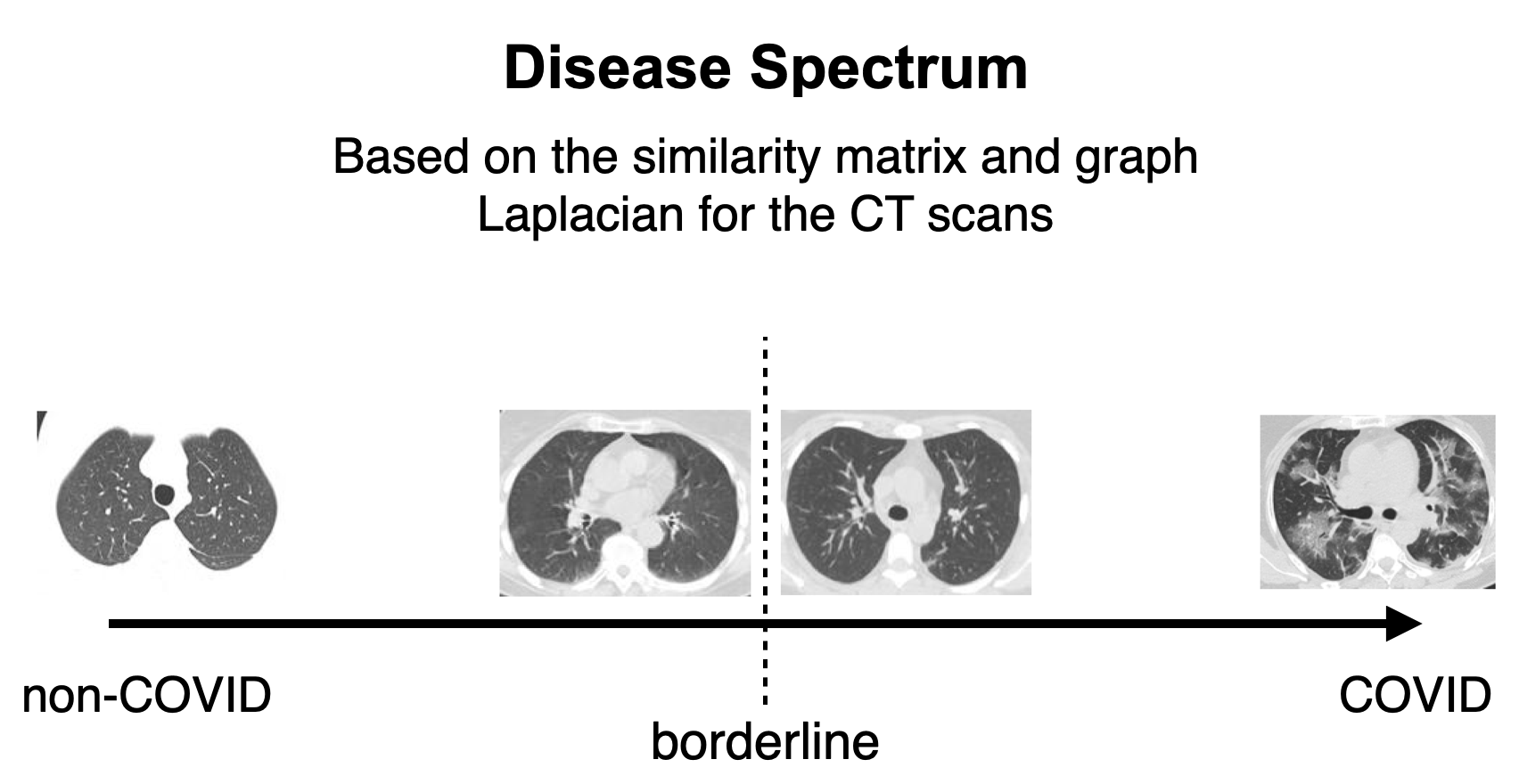}
   \caption{Disease spectrum inferred from labeled images of SARS-COV-2 CT-Scans.
   }
  \label{fig:covid_spectrum}
\end{figure}

Here, we leverage the similarities and dissimilarities among images to place them on a disease spectrum representing the severity of disease. The idea is to analyze the similarities of images from two different classes to automatically infer the disease spectrum. Figure~\ref{fig:covid_spectrum} shows the disease spectrum we infer for a dataset of SARS-COV-2 CT-Scans \citep{angelov2020explainable} containing 1252 images that are COVID-positive and 1230 images that are COVID-negative. Figure~\ref{fig:covid_samples} shows samples of images in this dataset. Moreover, Figures~\ref{fig:covid_isolated} to \ref{fig:covid_spec_not_ill} demonstrate specific images on different locations of this spectrum.


This disease spectrum has a borderline in the middle separating CT-scans of patients with COVID from healthy patients. Images of each class that are similar to images of other class will appear near the borderline. Images in one class that have large similarities to images in the other class would be considered in the middle of the disease spectrum, while images that have strong in-class similarities and weak out-class similarities would be placed away from the mid-spectrum, i.e., the borderline. This can be considered an unsupervised approach on labeled images with the aim to extract extra information from them. Labels define which patients have COVID-19, but they do not reveal how severe the disease is.


{\bf Interpreting the disease spectrum.} Result of this analysis can be used to identify images that are likely to be true-positives (highly infected) and true-negatives (non-infected), and also images that are near the borderline with respect to the disease. Images at the borderline will be highly useful for learning purposes by humans and machines. Humans would be able to analyze the minute differences between CT-Scans of infected and not-infected patients, and learn about the early signs of the disease. It would also be essential for predictive models to distinguish such differences and be able to detect which patients may be infected. Patients whose CT-Scans appear at the far right of the spectrum would be highly ill, with visible symptoms. It would be more useful if a model can correctly classify images that appear at the borderline which could be patients with mild or no apparent symptoms.

\section{Conclusions}

We considered a practical setting where a large dataset of medical images is gathered in a medical institution and we need to detect communities of similar images in order to proceed with classifying/labeling them. Our method uses wavelet decomposition of images in tandem with spectral methods to estimate the number of communities in dataset and to extract those communities. We experimented on two datasets of images related to SARS-COVID-19 and also a dataset related to histological images of colorectal cancer.

Our algorithm has implications for both unsupervised and supervised learning of medical images. For unsupervised learning, it facilitates the detection of communities of similar images in medical image datasets, facilitating the expensive process of labeling raw datasets. For supervised learning, our method can help in understanding fine-level similarities within each class and across classes. Such fine-level similarities can be used for training tasks such as triplet mining. Identifying images at the borderline of classes and flagging them for further review by medical experts may reduce the false predictions of deep learning models and make the automated process more reliable. Finally, we showed that analyzing the similarities may be used to infer a disease spectrum representing the severity of disease.

\bibliographystyle{splncs04nat}
\bibliography{refs}

\clearpage

\appendix

\setcounter{figure}{0}
\renewcommand{\thefigure}{A-\arabic{figure}}


\begin{figure}[h!]
  \centering
  \includegraphics[width=0.35\linewidth]{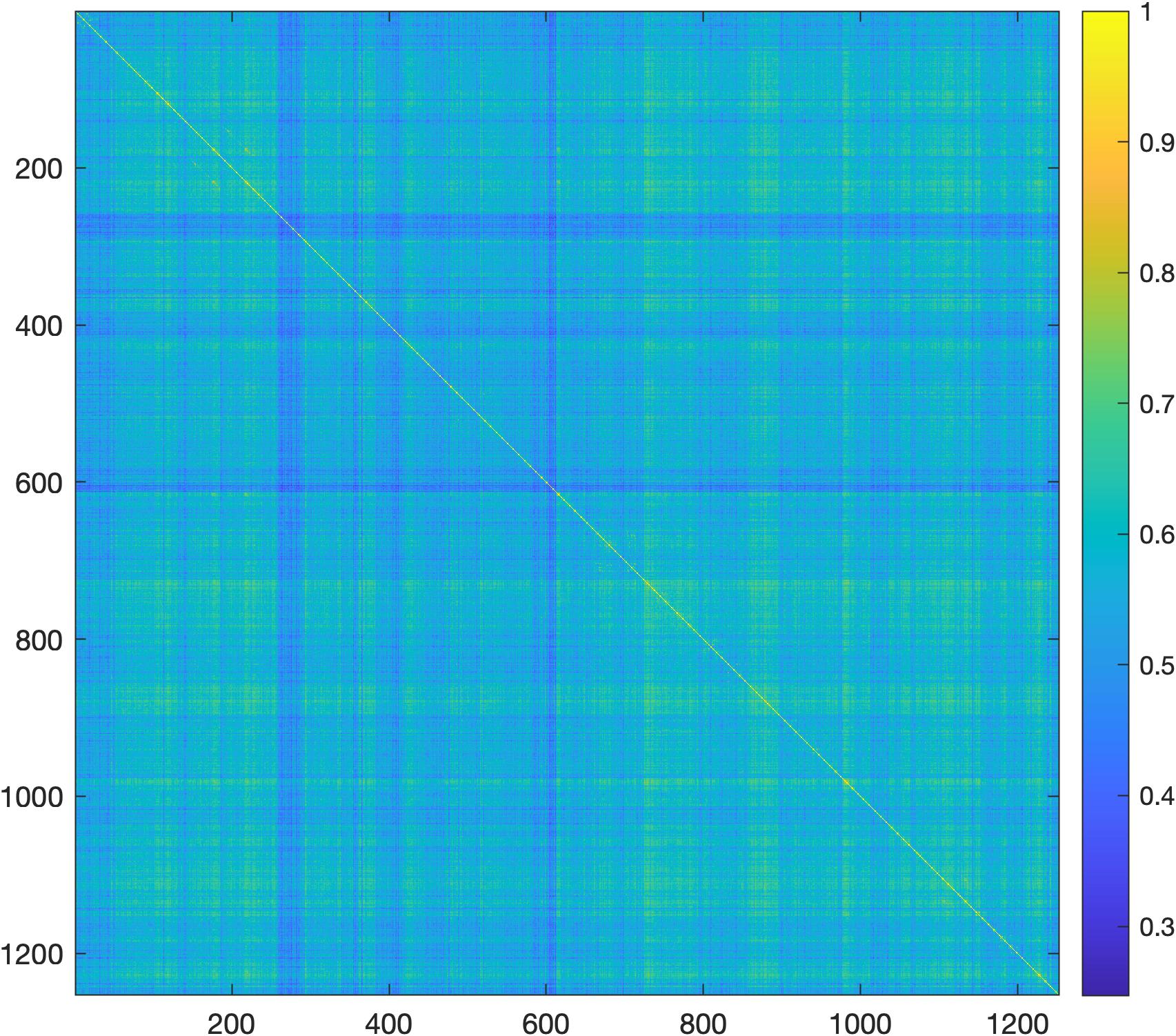}
  \caption{Similarity matrix of images in the COVID class.}
  \label{fig:covid_similaritym}
\end{figure}

\begin{figure}[h]
  \centering
  \fbox{
  \includegraphics[width=0.15\linewidth]{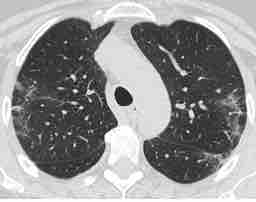}
  \includegraphics[width=0.15\linewidth]{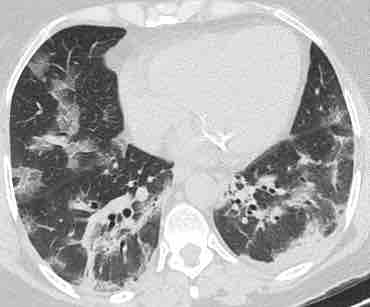}}
  \hfil
  \fbox{
  \includegraphics[width=0.13\linewidth]{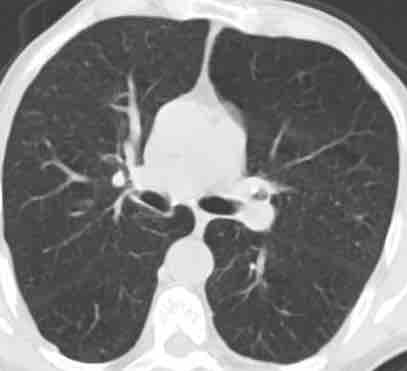}
  \includegraphics[width=0.13\linewidth]{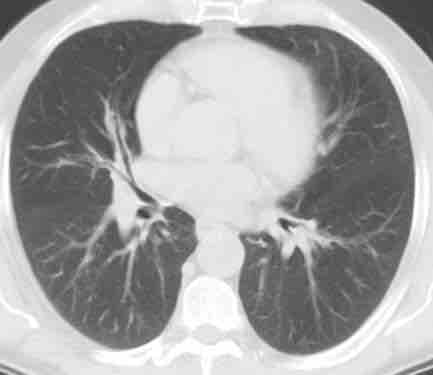}}
  \caption{Sample of images in the SARS-COV-2 CT-Scan dataset. Two images in the left box are from {\em infected} patients and the ones in the right box are from {\em non-infected} patients.}
  \label{fig:covid_samples}
\end{figure}

\begin{figure}[h!]
  \centering
  \includegraphics[width=0.15\linewidth]{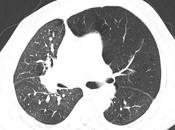}
  \includegraphics[width=0.15\linewidth]{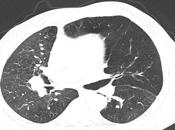}
  \includegraphics[width=0.15\linewidth]{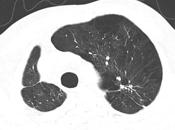}
  \includegraphics[width=0.15\linewidth]{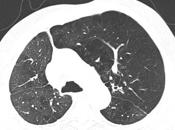}
  \caption{Isolated images in the COVID class, based on the similarity matrix in Figure~\ref{fig:covid_similaritym}.}
  \label{fig:covid_isolated}
\end{figure}

\begin{figure}[h!]
  \centering
  \includegraphics[width=0.15\linewidth]{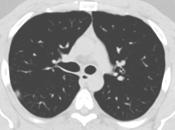}
  \includegraphics[width=0.15\linewidth]{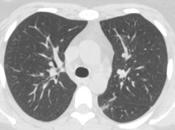}
  \includegraphics[width=0.15\linewidth]{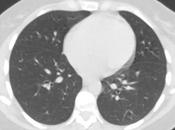}
  \includegraphics[width=0.15\linewidth]{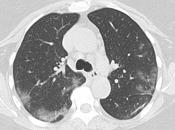}
  \caption{Borderline images: Images in the COVID class that are similar to images in the non-COVID class.}
  \label{fig:covid_spec_border_ill}
\end{figure}

\begin{figure}[h!]
  \centering
  \includegraphics[width=0.15\linewidth]{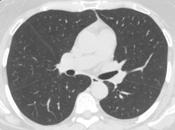}
  \includegraphics[width=0.15\linewidth]{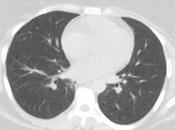}
  \includegraphics[width=0.15\linewidth]{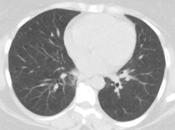}
  \includegraphics[width=0.15\linewidth]{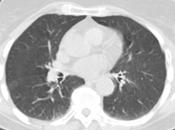}
  \caption{Borderline images: Images in the non-COVID class that are most similar to images in the COVID class.}
  \label{fig:covid_spec_border_not_ill}
\end{figure}

\begin{figure}[h!]
  \centering
  \includegraphics[width=0.15\linewidth]{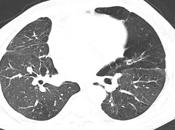}
  \includegraphics[width=0.15\linewidth]{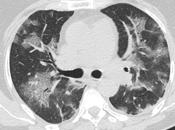}
  \includegraphics[width=0.15\linewidth]{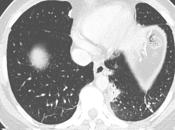}
  \includegraphics[width=0.15\linewidth]{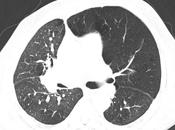}
  \caption{COVID images that are most dissimilar to non-COVID images. These may correspond to the infected side of the spectrum, far from the borderline.}
  \label{fig:covid_spec_ill}
\end{figure}

\begin{figure}[h!]
  \centering
  \includegraphics[width=0.15\linewidth]{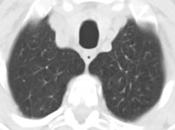}
  \includegraphics[width=0.15\linewidth]{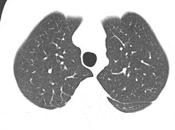}
  \includegraphics[width=0.15\linewidth]{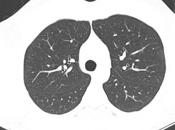}
  \includegraphics[width=0.15\linewidth]{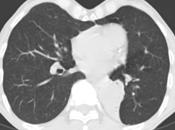}
  \caption{Non-COVID images that are most dissimilar to COVID images. These may correspond to the non-infected side of the spectrum.}
  \label{fig:covid_spec_not_ill}
\end{figure}

\clearpage

\end{document}